# Comparative study of type-II superconducting properties in polycrystalline NdFeAsO$_{0.88}$F$_{0.12}$ prepared by different methods

Y. Ding[1], Y. Sun[1], X. D. Wang[1], H. C. Wang[1], Z. X. Shi[1], Z. A. Ren[2], J. Yang[2], W. Lu[2]

1. Department of Physics, Southeast University, Nanjing, China 211189

Corresponding author: zxshi@seu.edu.cn

Mail address: dingyiseu@gmail.com

2. Institute of Physics and Beijing National Laboratory for Condensed Matter Physics, Chinese Academy of Sciences, Beijing 100190, P. R. China

Abstract

Polycrystalline NdFeAsO$_{0.88}$F$_{0.12}$ superconductors prepared by high pressure (HP) and ambient pressure (AP) methods were comparatively studied by magnetization and transport measurements. Upper critical field $H_{c2}$, irreversibility field $H_{irr}$ and the anisotropy parameter $\Gamma$ were estimated from resistance transition curves. The broadening of transition width was observed, and was ascribed to both $H_{c2}$ anisotropy and superconductivity inhomogeneity of samples. Magnetic hysteresis loops (MHLs) in low fields were measured to detect the trace of weak-link behavior. The reclosed hysteresis loops in low fields indicate that there are weak links in both samples. Magnetization critical current density $J_{cm}$ were derived from MHLs. Sample HP shows higher $J_{cm}$ than sample AP. Direct transport $I$-$V$ measurements show that the transport critical current density $J_{ct}$ are very low but persist up to 9 Tesla, suggesting intrinsic strong-link existing in both samples.

## 1. Introduction

The recent discovery of superconductivity in quaternary compound LaFeAsO$_{1-x}$F$_x$ [1] has introduced a new family of rare-earth iron oxypnictides superconductors with the general formula REFeAsO$_{1-x}$F$_x$. The transition temperature is above 50 K when RE=Nd [2] or Sm [3] and the upper critical field $H_{c2}$(0 K) exceeds 100 Tesla [4], suggesting promising potential applications. Several groups have reported the fabrications of superconducting REFeAsO$_{1-x}$F$_x$ wires [5-8]. However, these materials are significantly anisotropic [9] due to the layered structure, and have a low carrier density in the order of $10^{21}$ cm$^{-3}$ [10]. High $H_{c2}$ also implies a very short coherence lengths $\xi$. The similarity of basic superconducting properties to the cuprates indicates the possible existence of weak links and electromagnetically granular behavior, as already investigated by magnetization measurements and magneto-optic imaging in previous reports [11-13]. In this paper, type-II superconducting properties of samples prepared with both high pressure (sample HP) and ambient pressure (sample AP) methods were investigated systematically. Magnetization critical current $J_{cm}$ and transport critical current $J_{ct}$ were measured and compared directly. Related mechanisms were briefly discussed based on the comparative study between two samples.

## 2. Experimental

The superconducting NdFeAsO$_{1-x}$F$_x$ samples were prepared by both high pressure and ambient pressure methods. Preparation of sample HP is similar to reference [2]. The starting materials were mixed according to the nominal stoichiometric ratio of $NdFeAsO_{0.88}F_{0.12}$, then ground thoroughly and pressed into pellets. The pellets were sealed in boron nitride crucibles and sintered in a high pressure synthesis apparatus under 6 Gpa and the temperature of 1250 ℃ for 2 hours. Sample AP was prepared by the two-step solid-state reaction method. The stoichiometric mixture of starting materials was ground thoroughly and pressed into pellets, then sintered in an evacuated quartz tube at 1150℃ for 45 hours. Details can be found in [14].

Sample HP sized $1.16 \times 1.88 \times 4.5$ mm$^3$, weights 58 mg and is about 82% of the theoretical density (7.21 g/cm$^3$). Sample AP sized $0.62 \times 3.18 \times 4.66$ mm$^3$ weights 35 mg and is 53% dense. The crystal structure was characterized by powder X-ray diffraction (XRD) on an MXP18A-HF–type diffractometer with Cu-Kα radiation from 20° to 80° with a step of 0.01°. Microstructural and compositional investigations were performed using a Scanning Electron Microscope (SEM) equipped with an Energy Dispersive X-Ray Spectroscope (EDX). Transport and magnetization measurements were performed using a Quantum Design PPMS. Transport superconducting transitions were measured by using the conventional four-probe technique. During $I$-$V$ measurements, magnetic field was applied perpendicular to the current. The distance between two voltage probes was 1.06 mm for sample HP and 1.2 mm for sample AP. Transport critical current $J_{ct}$ was obtained from $I$-$V$ characteristic curves by choosing a criterion of 1 μV/cm. DC magnetization measurements was performed using the VSM option of the PPMS with magnetic field applied along the longest dimension of the samples. Magnetic hysteresis loops (MHLs) in low fields were measured with the sensitivity of $10^{-5} \sim 10^{-6}$. Magnetic critical current densities $J_{cm}$ were estimated using the Bean model from the MHLs.

## 3. Results and discussion

*3.1* XRD and SEM

Figure 1 shows the XRD pattern of sample HP and AP. Almost all the diffraction peaks can be well indexed on the basis of tetragonal ZrCuSiAs-type structure with a P4/nmm space group, confirming the main phases are $NdFeAsO_{1-x}F_x$. The calculations for the crystal parameters were performed on the basis of a least-squares fit using the Jade programs. The a-axis and c-axis values are 0.3969 nm and 0.8531 nm for sample HP, while a=0.3974 nm and c=0.8576 nm for sample AP. Sample HP has smaller lattice parameters than sample AP, suggesting more F doped into sample HP. The average grain size calculated by using Scherrer formula is about 64.2 nm for sample HP and 92.2 nm for sample AP. Figure. 2 (a) and (b) are the SEM images of sample HP and AP after metallographic preparation. Compared with sample AP, sample HP shows a denser structure constituted of rectangular shaped tabular crystals. Sample AP is porous, and only some parts of the region are well connected. No large cracks are observed in both samples. Figure 2 (c) and (d) are the SEM images of HP and AP with higher magnification of 50000×. Sample HP shows better connectivity between grains than AP. It is also noted that sample HP has more fine grains, which is consistent with the calculation. The reason is that a long time sintering up to 45 h promotes grain coarsening in sample AP. The EDX analysis indicated that for both samples the composition of Nd:Fe:As on the crystals is close to 1:1:1. But the exact composition of O and F cannot be determined due to the light masses.

*3.2* Upper critical field and anisotropy

Figure 3 shows the temperature dependence of resistivity with applied fields up to 9 T. Under zero fields, onset transition temperature of sample HP and AP determined by 90% normal-state resistivity are about 51 K and 44 K, respectively. Sample HP has higher $T_c$ because high-pressure synthesis method promotes F substitution for O [14], which is consistent with the XRD results. Figure 4 shows the temperature dependence of three characteristic fields $H_{90}$, $H_{10}$ and $H_2$ at which the resistivity reaches 90%, 10% and 2% of the value of normal-state. $H_{90}$ is associated with $H_{c2}^{ab}$ because grains with their a-b planes parallel to the applied field become superconducting first upon cooling. According to a previous analysis for polycrystalline samples [4], we assume that, because of the strong angular dependence of $H_{c2}(\theta)$, $H_{10}(T)$ scales like the upper critical field parallel to the c-axis of grains [15], i.e. $H_{c2}^{c}$. $H_2$ is close to $H_{irr}$. $H_{c2}^{ab}$ and $H_{c2}^{c}$ results of single crystal [9] and polycrystalline samples in other reports [15, 16] are included in Figure 4 for direct comparison. Similar slopes of $H_{c2}^{ab}$ and $H_{c2}^{c}$ can be observed among different samples. Taking the average slope $-\dfrac{\mathrm{d}H_{c2}^{ab}}{\mathrm{d}T}\Big|_{T_c} = 8$ T/K for sample HP and 6.4 T/K for sample AP, the single band Werthamer-Helfand-Hohenberg (WHH) formula $H_{c2}(0) = -0.693 T_c \dfrac{\mathrm{d}H_{c2}}{\mathrm{d}T}\Big|_{T_c}$ [17] yields $H_{c2}^{HP}(0) = 283$ T and $H_{c2}^{AP}(0) = 195$ T, which clearly exceed the Pauli limit. Since the high-field Pauli-limiting behavior was observed in the iron-based superconductors [18], the spin paramagnetic and spin-orbit effects should be taken into account when estimating $H_{c2}(0)$. In Figure 5, $H_{90}(T)$ and $H_{10}(T)$ was normalized to the respective values of $T_c^2$ for each sample and plotted as functions of the reduced temperature $t = T/T_c$. $H_{10}(T)$ data collapse onto single curve while $H_{90}(T)$ do not. The reason is that, as investigated by Jaroszynski et al. [15], the RE-1111 (RE=La, Sm and Nd) polycrystalline superconductors are likely in the clean limit, $\xi_a$ and $\xi_c < l$, where $\xi_a$ and $\xi_c$ are coherence lengths perpendicular to the c axis and along the c axis, respectively, $l$ is the mean-free path. In clean limit, $\xi_a \propto \xi_c \propto 1/T_c$, yields $H_{c2}^c \sim \phi_0 / 2\pi\xi_a^2 \sim T_c^2$, which is consistent with the scaling of $H_{10}(T)$ curves. $H_{90}$ does not follow the scaling because $H_{c2}^{ab} \sim \phi_0 / 2\pi\xi_a\xi_c = \phi_0 / 2\pi\xi_a^2\Gamma^{-1/2}$, where $\Gamma = \rho_c / \rho_{ab} = \xi_{ab}^2 / \xi_c^2$ is the effective mass or resistivity anisotropy parameter, which may be temperature dependent [19] and different between sample HP and AP. Since $\xi_a$ is proportional to $1/T_c$, the slope of $H_{c2}^{ab}$, $\dfrac{dH_{c2}^{ab}}{dT}$ near $T_c$ is proportional to $T_c\Gamma^{1/2}$. By choosing $-\dfrac{dH_{c2}^{ab}}{dT}\bigg|_{T_c} = 8$ T/K for sample HP, 6.4 T/K for sample AP and $\Gamma = 15$ [20], $T_c = 28$ K, $H_{c2}^{ab}$ slope of 2.7 T/K for LaFeAs(O,F) [15], $\Gamma_{HP} \sim 15(8T_c^{La} / 2.7T_c^{HP})^2$ is estimated to be about 40 and $\Gamma_{AP} \sim 15(6.4T_c^{La} / 2.7T_c^{AP})^2 \sim 34$. So obtained $\Gamma$ of sample HP and AP are close to the results of single crystal [9], that $\Gamma^{1/2} \leq 6$. The difference between $\Gamma_{HP}$ and $\Gamma_{AP}$ is caused by the less doping level in sample AP, which lead to a reduction of the interplanar coupling that suppresses both $T_c$ and $dH_{c2} / dT$ [21]. Additionally, defects and vacancies caused by F substitution of O mainly shrinks c-axis parameter [22], suggesting that the electron mean free path along c-axis $l_c$ decreases upon doping. And in plate-like crystals, $l_c$ is also limited by the grain dimension along c-axis. Sample HP has a higher doping level, and has more fine grains than sample AP as shown in Figure 2 (c) and (d), indicating a reduced $l_c^{HP}$. According to the equation $\dfrac{1}{\xi_c(l)} = \dfrac{1}{\xi_0} + \dfrac{1}{\alpha l_c}$, where $\xi_0$ is the coherent length of pure superconductor, $\alpha$ is a constant with the order of magnitude of 1, $\xi_c$ decreases with decreasing $l_c$. Thus the reduced $l_c^{HP}$ leads to a shorter $\xi_c^{HP}$ and a larger $\Gamma = \dfrac{\xi_a^2}{\xi_c^2}$ value.

*3.3* Superconducting transition width

Two fractions of transition width were defined as $\Delta T_c = T(H_{90}) - T(H_{10})$ and $\Delta T_{irr} = T(H_{10}) - T(H_2)$. The field dependence of $\Delta T_c$ and $\Delta T_{irr}$ is shown in Figure 6. In zero field, $\Delta T_c$ is caused by $T_c$ distribution among grains due to sample inhomogeneity. Under applied fields, $\Delta T_c$ is further influenced by anisotropy of grain superconductivity and connectivity, while $\Delta T_{irr}$ is mainly broadened by flux flow and flux creep. With fields applied on, $\Delta T_c(H)$ shiftes to higher values. In lower fields, the curves show steep initial slopes, because worsen connectivity due to rapidly suppression of weak links postponed the transition, and the apparent anisotropy $\Gamma^*$ was enhanced to $\sim \dfrac{H_{c2}^{ab*}}{H_{c2}^{c*}}$, due to distribution of upper critical fields caused by inhomogeneity, where $H_{c2}^{ab*}$ is the highest upper critical field and $H_{c2}^{c*}$ is the lowest lower critical field of the grains. The transition width broadening caused by anisotropy was proposed by Eisterer [23] as $\Delta T_a = \dfrac{\sqrt{(\Gamma-1)p_c^2+1}-1}{(-\partial H_{c2}/\partial T)}H_e$, where $p_c$ is the percolation threshold. When the probability $p$, that a grain is superconducting, becomes equal to $p_c$, the first continuous superconducting current path occurs and the resistivity disappears for sufficiently small currents. $H_e$ is the magnetic field, and assuming the $dH_{c2}/dT$ and $\Gamma$ are constant within the transition region near $T_c$. In high fields, weak links were fully suppressed so $\Delta T_c$ can be described by $\Delta T_a + \Delta T_{in}$, where $\Delta T_{in}$ represents the broadening of the transition due to a distribution of $T_c$ of grains. Since the slopes of $H_{c2}^c(T)$ curves seem comparable among samples with different $T_c$ as observed in Figure 4, $\Delta T_{in}$ is nearly independent of fields.

In Figure 6, taking $\Delta T_{in}$ and $p_c$ as fitting parameters, $\Delta T_{in}^{HP} = 4.74$, $\Delta T_{in}^{AP} = 3.52$, $p_c=0.7$, and $\Gamma_{HP}=40$, $\Gamma_{AP}=32$, $H_{c2}$ slope of $-8$ T/K for sample HP and $-6.4$ T/K for sample AP, the calculations fit the experimental data well in the high fields. $\Delta T_{in}^{HP}$ is higher than $\Delta T_{in}^{AP}$, suggesting sample HP is less homogeneous than sample AP. In contrast, $\Delta T_{irr}$ of sample AP is higher than sample HP as shown in the inset of Figure 6, which indicates stronger flux creeping and flow in sample AP. Our results suggest that high-pressure short time synthesis method may not improve sample homogeneity, but may be effective on inducing vortex pinning centers such as intragrain, atomic-scale defects, which are reduced by long time sintering.

### *3.4* Weak-link behavior

The temperature dependence of Magnetic moment $\chi(T)$ was measured in zero-field cooled (ZFC) and field cooled (FC) states in the range of 5-60 K under an external field of 15 Oe. Geometric effect was considered by using an effective demagnetization factor proposed by Brandt [24]. Superconducting fraction was estimated from ZFC and FC $\chi(T)$ curves to be 78% for sample HP and 40% for sample AP. Magnetization measurements in low field were performed to investigate the weak-link behavior. Before measuring each loop, residual field was reduced to a value less than 1 Oe and the sample was warmed up to a temperature sufficiently higher than $T_c$ to fully expel the flux trapped inside. During the measurements, the magnetic field was increased from 0 Oe to a maximum, $H_{max}$, then decreased to 0 Oe. The sweeping rate was 12 Oe/sec, and the data were collected one point per Oe. A series of loops were measured using monotonic increased $H_{max}$. Figure 7 shows the low-field magnetization curves with several $H_{max}$ of sample HP at 11 K, 34 K and of sample AP at 10 K, 30 K, i.e., normalized temperature $T/T_c$ of 0.22 and 0.67. After demagnetization correction, the slopes of linear reversible curves are close to -0.78 and -0.40 for sample HP and AP, respectively.

As shown in Figure 7 (a), (b) and (d), the hysteresis loops opened in low fields but reclosed between 57-60 Oe, 13.3-15 Oe and 7.5-10 Oe, respectively. This can be explained by the existence of weak-link boundaries. For granular superconductors, the weak superconductivity in grain boundaries is sensitive to magnetic field thus the entering field of grain boundaries $H_{en}^w$ is less than $H_{en}$ of grains. When applied field $H_{max}$ exceeded $H_{en}^w$, a first hysteresis loop opened owing to trapping of intergrain vortices. At larger field, the hysteresis loop closed (within the sensitivity of the magnetometer) and a second reversible magnetization curve appeared because the superconductivity of weak links was fully suppressed while all grains may still be in the Meissner state. Further increasing $H_{max}$, the hysteresis loop opened again due to entrance of vortices into grains. In Figure 7 (a), (b) and (d), the second reversible magnetization range is short, indicating that $H_{en}^w$ is close to $H_{en}$ in both samples. In the case of Fig 7 (c), low field loops of sample AP at 10 K were broadened gradually with increasing $H_{max}$ and no second reversible curve appeared. It may be attributed to a distribution of $H_{en}^w$ and $H_{en}$ of grains, which results in an overlap of $H_{en}^w$ and $H_{en}$, and an obscured $H_{c2}^{w*}$, at which field weak links are totally broken. Recently, weak-link behaviour of iron-based superconductor was also detected by the harmonics of the ac magnetic susceptibility [25].

*3.5* Magnetization and transport $J_c$

To further investigate and compare the superconducting properties of the two samples, magnetization and transport measurements were performed. $J_{cm}$ was extracted from MHLs using the Bean model, $J_c = 20\Delta M/a(1-a/3b)$, where $\Delta M$ is the width of the MHLs, $a$ and $b$ ($a<b$) are thickness and width of the sample. As shown in Figure 8, the $J_{cm}$ values show strong field dependence in low field region. And sample HP has higher $J_{cm}$ than sample AP. Since the normalized temperature $T/T_c$ at which the measurements performed is quite close for each sample, much higher $J_{cm}$ in sample HP is not likely caused by different $T_c$, but due to stronger flux pinning and better global connectivity. Yamamoto et al. [13] pointed that there are distinct scales of current flowing in the Nd-1111 superconductors, so the MHLs width $\Delta M$, from which $J_{cm}$ was derived can be written as $\Delta M^{\text{global}} + \Delta M^{\text{grain}}$, where $\Delta M^{\text{global}}$ is the contribution of bulk current flowing over the whole sample and $\Delta M^{\text{grain}}$ denotes the fraction caused by intra-grain currents. Sample HP is better connected than porous sample AP as shown in Figure 2, and has stronger flux pinning as discussed in 3.2, leading to higher $\Delta M_{\text{HP}}^{\text{global}}$ and $\Delta M_{\text{HP}}^{\text{grain}}$. With increasing fields, intergrain connectivity is suppressed and supercurrent is confined to circulating within smaller better-connected regions such as conglomerate particles [11]. The loop size reduces, resulting decrease of $\Delta M^{\text{global}}$. In the high field region, $\Delta M_{\text{HP}}^{\text{grain}}$, which is caused by current loops in superconducting grains with very high $H_{\text{irr}}$, mainly contribute to $\Delta M$, thus the derived $J_{cm}$ values is much smaller than real $J_{cm}$ in grains and show weak field dependence.

$J_{ct}$ was obtained from $I$-$V$ curves measured at 5, 10, 20 and 30 K. Figure 9 shows electric field versus $J_{ct}$ curves at 20 K in fields of 0-90 KOe for sample HP and 0-10 KOe for sample AP. Steeper transition of sample HP than AP may be caused by different effect pinning barrier, which will be investigated elsewhere. Figure 10 shows $J_{ct}$ of sample HP and AP at various temperatures under magnetic fields up to 1 T. $J_{ct}$ values of sample HP at 20 K up to 9 T are also presented. $J_{ct}$ is 10-100 times lower than $J_{cm}$, because in transport measurements, supercurrents can go only through the region between electrodes. Thus $J_{ct}$ is determined by pinning of grains and also limited by bulk connectivity of this region. As shown in Fig. 10, $J_{ct}$ drop quickly from 0 to 0.1 T, indicating weak links broken by magnetic fields. Compared with $J_{ct}$ of Sm-1111 superconductors measured at 4.2 K [26] and 39 K [27], the self-field $J_{ct}$ of sample HP and AP are much lower, suggesting very small superconducting cross sections. This may be caused by multiple extrinsic factors such as macroscopic phase inhomogeneity [28], grain boundary cracks and impurities at grain boundaries [29], which can significantly reduce the current paths. However, $J_{ct}$ in high fields are close to $J_{ct}$ values in reference [26] and [27], indicating that there do exist some intrinsic strong links even in the sample prepared in ambient pressure. The NdFeAsO$_{1-x}$F$_x$ superconductors are still promising for application if phase purity can be improved and the amount of strong links can be increased.

## 4. Conclusions

In summary, superconducting properties of polycrystalline NdFeAsO$_{0.88}$F$_{0.12}$ superconductors prepared by high pressure and ambient pressure methods were investigated by magnetization and transport measurements. The comparison between two samples leads to the following conclusions. (1) Both samples show very high $H_{c2}$ and pronounced anisotropy. (2) Higher $J_{cm}$ in sample HP is due to enhanced intra-grain vortex pinning and improved local connectivity, which is related to the specific feature of the high-pressure synthesis. (3) Strong links do exist in both samples although $J_{ct}$ values are very low, which may be caused by weak links and very small superconducting cross section.

**Acknowledge**

We are very grateful to Prof. M. Sumption for discussions. This work was supported by the scientific research foundation of graduate school of Southeast University (Grant No. YBJJ0933), by the Jiangsu Industry Support Project (Grant No. BE2009053), by the Natural Science Foundation of China (NSFC, Grant No. 10804127), by the Cyanine-Project Foundation of Jiangsu Province of China (Grant No. 1107020060) and by the Foundation for Climax Talents Plan in Six-Big Fields of Jiangsu Province of China (Grant No. 1107020070).

S. Pace, Phys. Rev. B **78** (2008) 224523.

**Figure Captions**

Figure 1. X-ray powder diffraction patterns of sample HP and AP.

Figure 2. The SEM images of sample HP and AP with low and high magnifications. (a) HP, 2000×. (b) AP, 3000×. (c) HP, 50000×. (d) AP, 50000×.

Figure 3. Temperature dependence of resistivity under magnetic fields in the range of 1~9 T for sample HP (a) and AP (b). Superconducting transition from zero resistivity state to normal state is schematically shown in (b).

Figure 4. Characteristic fields $H_{90}$, $H_{10}$ and $H_2$ versus temperature for sample HP and AP. $H_{90}$, $H_{10}$ and $H_2$ was determined by 90%, 10% and 2% resistivity of normal-state, and is associated with $H_{c2}^{ab}$, $H_{c2}^{c}$ and $H_{irr}$, respectively. $H_{c2}^{ab}$ and $H_{c2}^{c}$ results of reference [9, 15, 16] are included for direct comparison.

Figure 5. $H_{90}$ and $H_{10}$ divided by $T_c^2$ are plotted as a function of reduced temperature $T/T_c$.

Figure 6. The field dependence of transition width defined as $\Delta T_c = T(H_{90})-T(H_{10})$. Solid and point lines are fitting results with parameters (see main text) for sample HP and AP, respectively. Inset shows transition width defined as $\Delta T_{irr}=T(H_{10})-T(H_2)$ versus magnetic fields.

Figure 7. Magnetic hysteresis loops in low field measured in normalized temperature $T/T_c$=0.22 and 0.67. (a) HP at 11 K. MHLs reclosed between 57 and 60 Oe. (b) HP at 34 K.

MHLs opened at 10 Oe and reclosed between 13.3 and 15 Oe. (c) AP at 10 K. MHLs broadened gradually with increasing $H_{max}$ and no second reversible curve appeared (d) AP at 30 K. MHLs reclosed between 7.5 and 10 Oe.

Figure 8. Comparison of $J_{cm}(H)$ between sample HP and AP up to 1 T.

Figure 9. Electric field versus $J_{ct}$ curves at 20 K (a) HP, 0-90 KOe. (b) AP, 0-10 KOe.

Figure 10. $J_{ct}(H)$ curves of sample HP and AP at 5, 10, 20 and 30 K. Inset shows the same plots displayed in semilogarithmic scale.

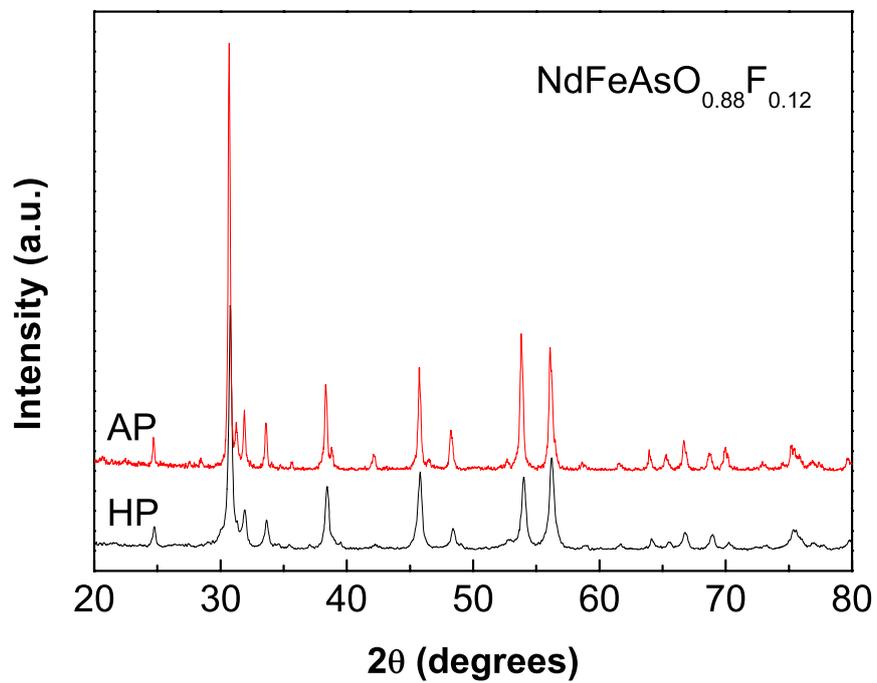

**Figure 1 Y. Ding *et al.***

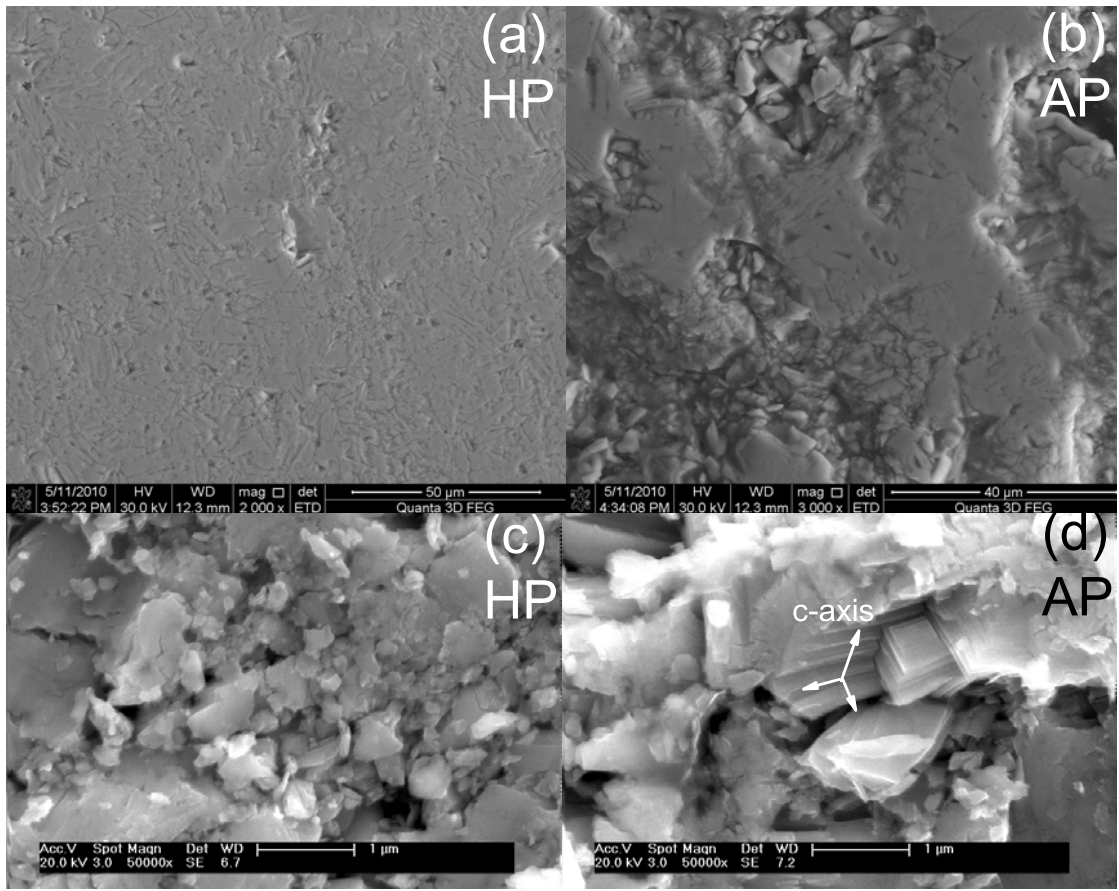

**Figure 2 Y. Ding *et al*.**

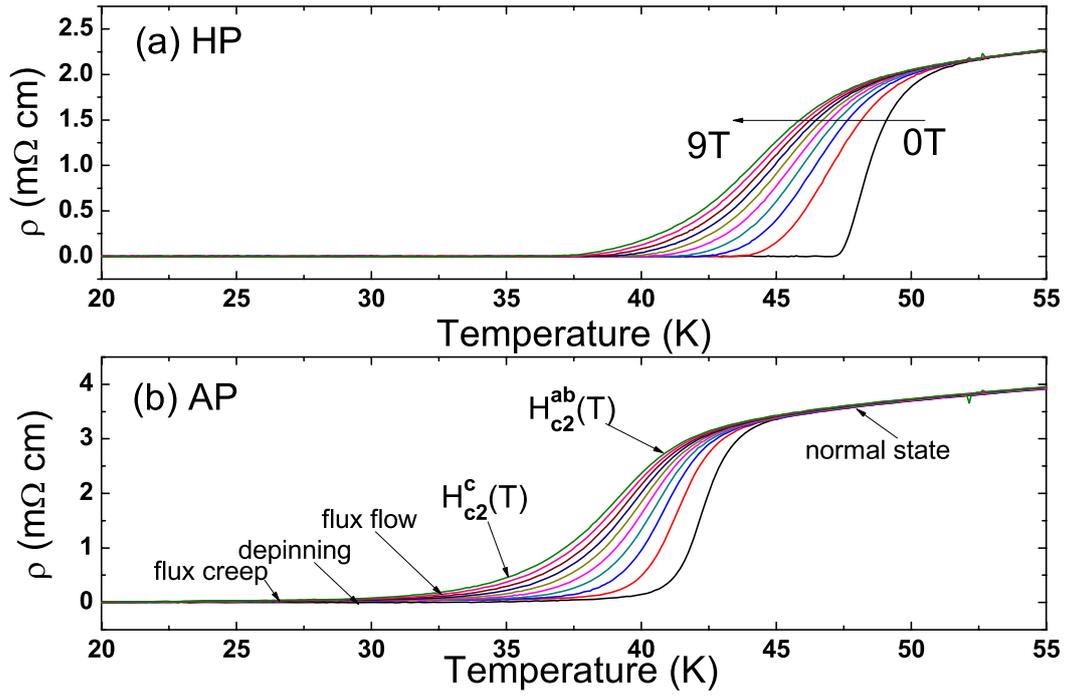

**Figure 3 Y. Ding *et al.***

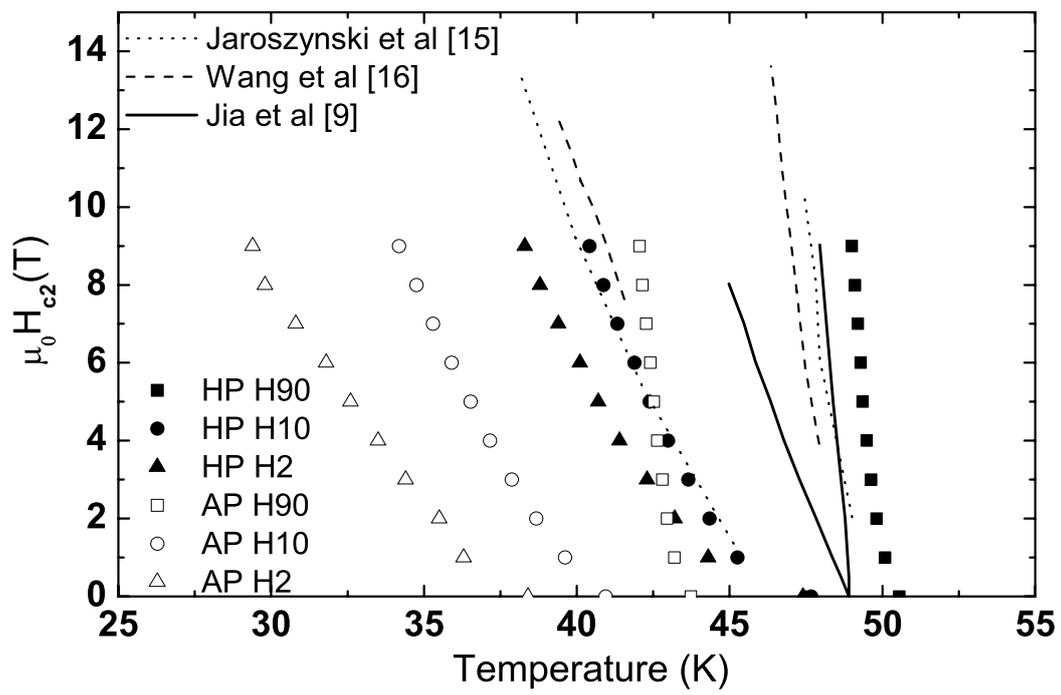

**Figure 4 Y. Ding *et al.***

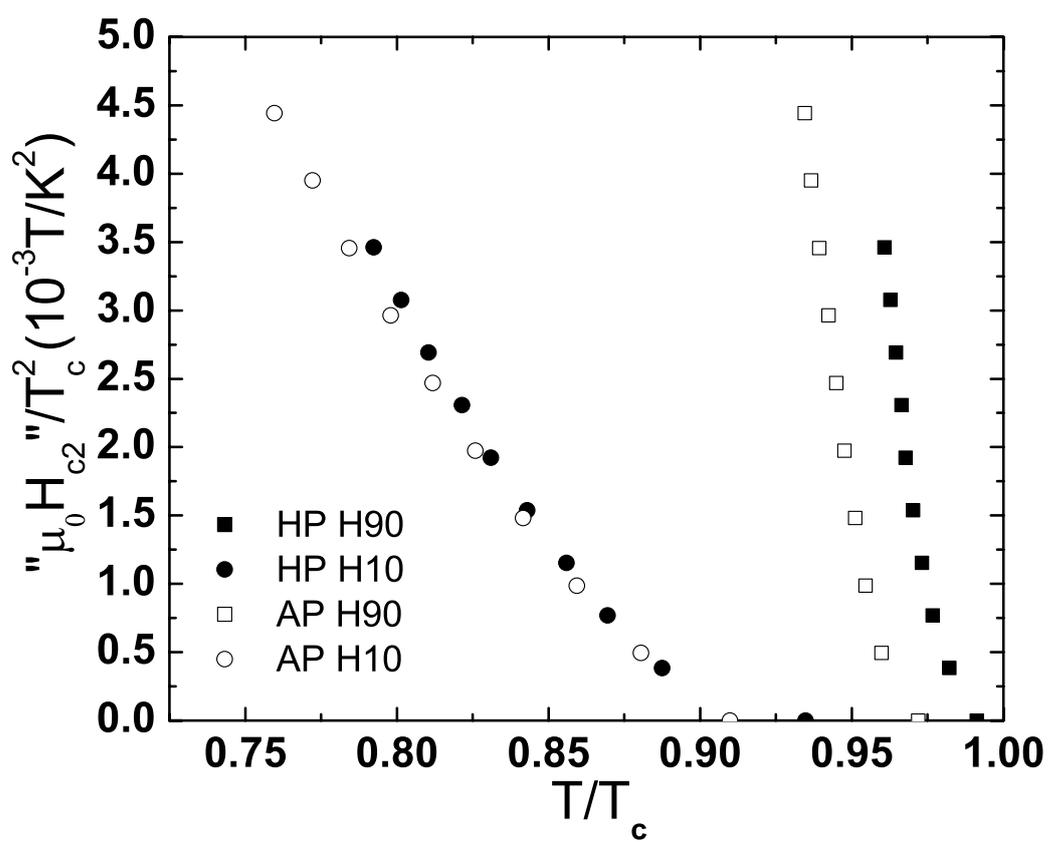

**Figure 5 Y. Ding *et al*.**

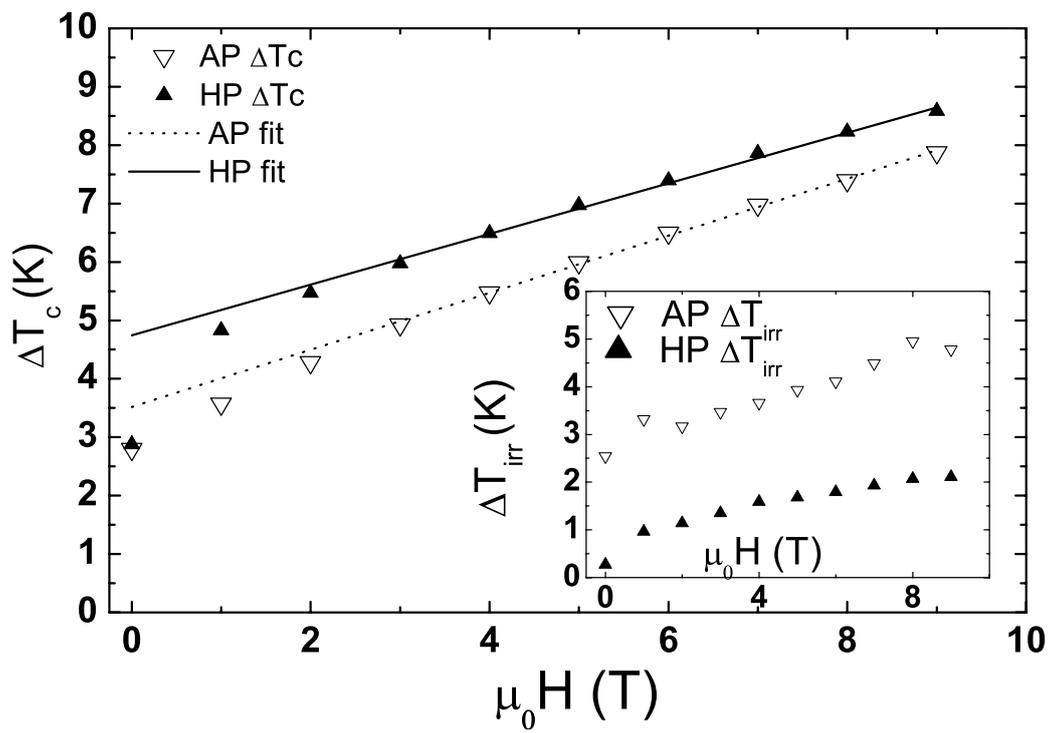

**Figure 6 Y. Ding** *et al.*

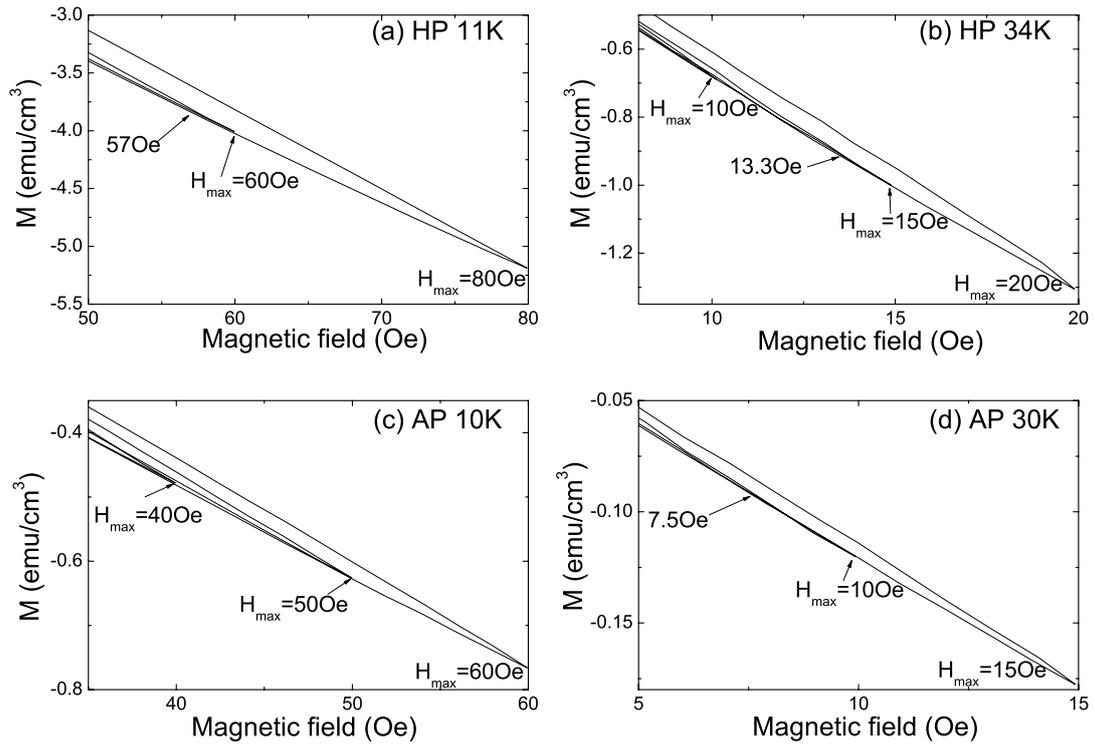

**Figure 7 Y. Ding *et al.***

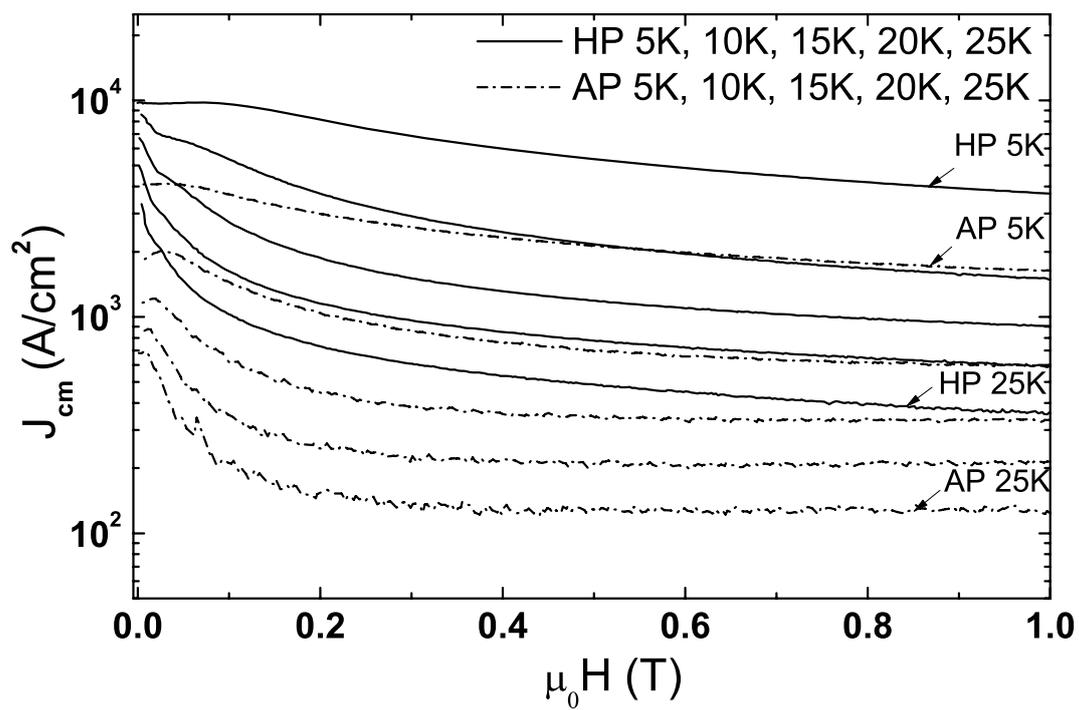

**Figure 8 Y. Ding** *et al.*

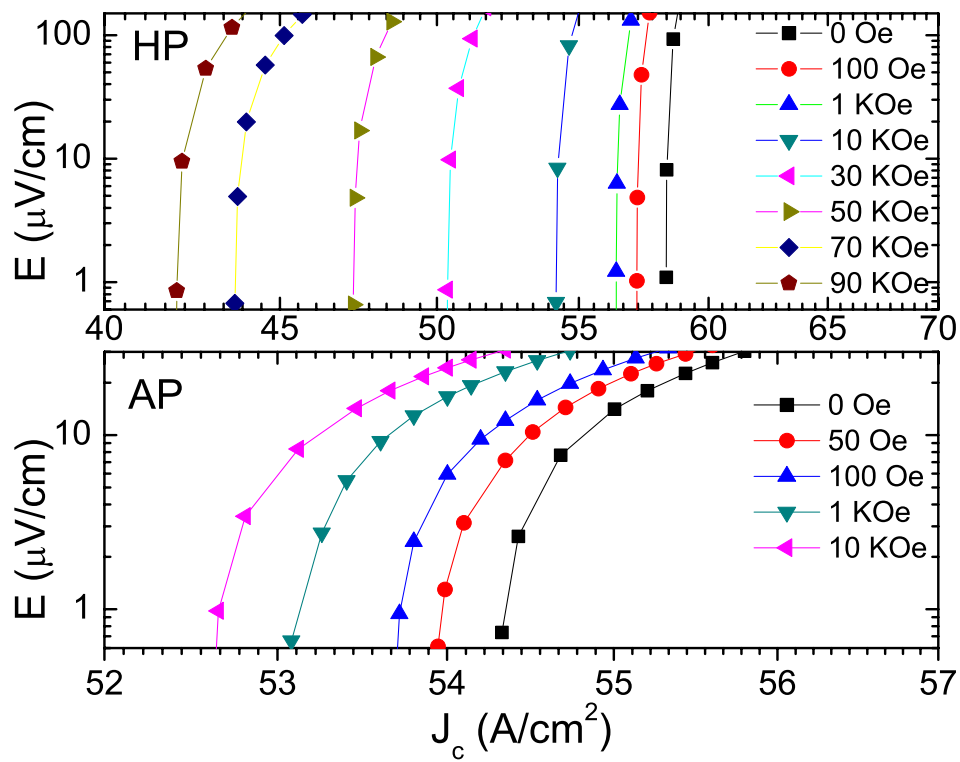

**Figure 9 Y. Ding** *et al.*

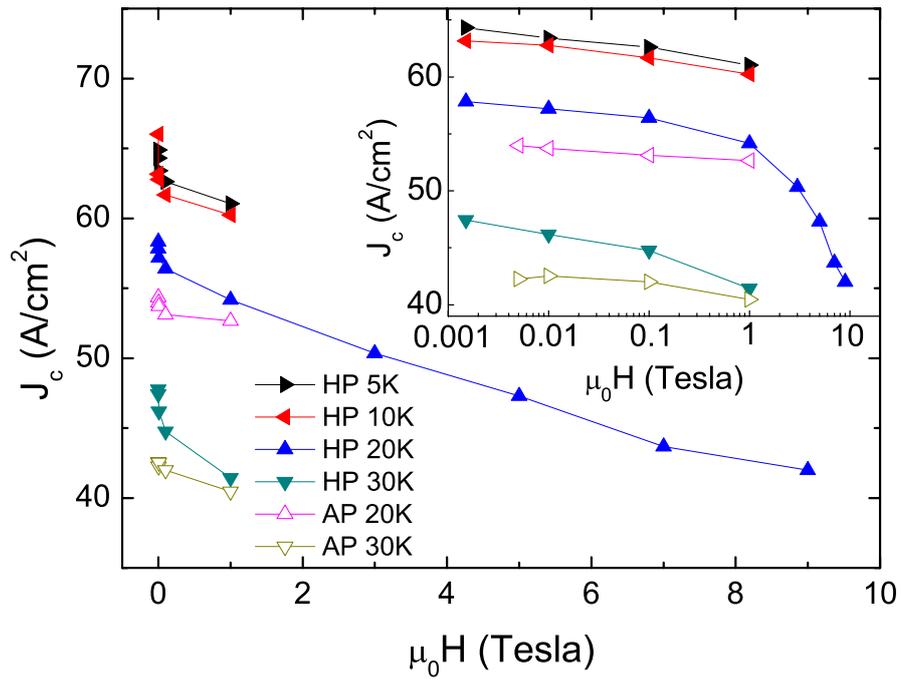

**Figure 10 Y. Ding** *et al.*